\documentclass[aps,prl,twocolumn,groupedaddress]{revtex4}

\usepackage{graphics}
\begin{document}


\title{Crossover between SC states in an unconventional superconductor UCoGe driven by ferromagnetic spin fluctuations}


\author{J. Poltierov\'{a} Vejpravov\'{a}, J. Posp\'{i}\v{s}il, and V. Sechovsk\'{y}}
\affiliation{Charles University in Prague, Faculty of Mathematics
and Physics, Department of Condensed Matter Physics, Ke Karlovu 5,
121 16 - Prague 2, Czech
Republic}
\email[J.P.V.:]{jana@mag.mff.cuni.cz}


\date{\today}

\begin{abstract}
UCoGe is reported as a weak ferromagnetic (FM) two-band
superconductor (SC) with the critical temperature,
$T_{\mathrm{SC}}\sim$ 0.7 K and the Curie temperature,
$T_{\mathrm{C}}\sim$ 3 K at ambient pressure. We report exotic
attributes of the SC and FM state in moderate magnetic fields. The
observed phenomena clearly demonstrate that the SC regime is
superior to the paramagnetic state in the vicinity of
$T_{\mathrm{SC}}$. Above $T_{\mathrm{SC}}$, the zero-field state
is characterized by a regime with strong FM spin fluctuations,
suggesting proximity of the FM quantum critical point (FM-QCP). In
addition, we observed that the robust SC regime develops
independently on the existence or lack of the long-range FM
ordering.
\end{abstract}


\maketitle

A delicate interplay between the quantum criticality and
superconductivity (SC), explicitly in the vicinity of the
ferromagnetic quantum critical point (FM-QCP) represents the
decisive factor in the magnetically driven superconductivity
\cite{1,2,3,4,2o,2t}. Recently, Huy \textit{et al.} reported on
UCoGe, which is typified as a weak ferromagnet undergoing a
subsequent transition into the coexistence of FM and SC below the
critical temperature, $T_{\mathrm{SC}}$ = 0.8 K \cite{5}. The
magnitude and anisotropy of the upper critical field suggest the
$p$-wave type of SC and points to an axial SC state with nodes
along the easy magnetization direction ($c$-axis), interpreted in
terms of an unusual two-band SC state \cite{6}. UCoGe is in fact
unique because of its extremely low $T_{\mathrm{C}}$ and small
spontaneous magnetic moment, indicating proximity of the FM
instability. The UCoGe case has been anticipated as the first
example of an unconventional SC (USC) stimulated by critical
fluctuations associated with a FM-QCP. In other words, the FM spin
fluctuations (FM-SF), sensitively tuned by an external magnetic
field serve as a coupled control parameter mediating the onset of
SC.

According to Doniach \cite{doniach}, the phonon-induced $s$-wave
SC in an exchange-enhanced transition metals is generally
suppressed by FM-SF, in the neighborhood of the $T_{\mathrm{C}}$.
In contrast, the FFLO theory \cite{fflo} revealed a SC state with
a spatially modulated order parameter (OP) coexisting with a
long-range FM order in metallic systems with the
magnetic-impurity-induced FM. The spin-correlation theory of the
FM state, based on the interactions mediated by SF between the
fermions proposes an enhancement of the pairing correlations
through the FM-SF \cite{1t}. The SC phase diagram based on this
approach comprises two SC phases. The $s$-wave type (generalized
FFLO) is established in the FM state, however the $p$-wave state
exists in the paramagnetic region on the border of the FM
instability, and is expected to vanish at the QCP. On the
microscopic scale, the critical temperature of the SC transition
depends on the difference of the pairing interaction and density
of states on the Fermi level, respectively, for the particles with
the opposite orientation of the corresponding pseudospins. In
principle, the competition of the stimulating and suppressing
effects on $T_{\mathrm{SC}}$  in FM-SCs determine phase diagrams
of these unique materials.

In this Letter, we report exotic features observed in low magnetic
fields on a series of polycrystalline and single-crystalline
samples of UCoGe. First, we inspected in detail the proposed
zero-field ferromagnetism below the $T_{\mathrm{C}}$. However, it
was not straightforwardly evidenced from our experimental results,
even on the annealed single-crystalline sample. Further, we
focused on the anomalous behavior of the critical SC temperature,
$T_{\mathrm{SC}}$ in moderate magnetic fields indicated by
detailed electrical resistivity and magnetoresistance
measurements.

The polycrystalline UCoGe samples were prepared by arc melting of
stoichiometric amounts of high-purity constituents (U - 2N, Co -
3N, Ge - 5N) in the stoichiometric ratio (1:1:1). The samples were
annealed at 800 and 900 $^{\circ}$C, respectively, for 10 days.
The reported single crystal of UCoGe has been isolated from a
large polycrystalline button as a plate of dimensions 2 x 1 x 0.5
mm. We want to point out, that the quality of this
single-crystalline grain was by far better then any of the
attempts to grow a crystal by Czochralski technique in a tri-arc
furnace. The single crystal was investigated as cast, and
additionally annealed at 900 $^{\circ}$C for 10 days. The phase
composition of the prepared materials was checked by microprobe
analysis and X-ray powder diffraction (XRD). The single crystal
was further oriented by Laue method in back-scattering geometry;
the proper crystallinity and orientation were carefully checked
from both sides of the sample plate. All samples (including the as
cast polycrystal with a considerable amount of foreign phases)
clearly manifest the transition to the SC state at
$T_{\mathrm{SC}}$, but the proposed FM features were much
disputable. The typical residual resistivity ratio ($RRR$) of
polycrystalline materials  reached 10; the $RRR$ of the
single-crystal (along the $c$ axis) was about 20. The XRD data can
be reliably described by two structure types CeCu$_2$ and TiNiSi,
represented by the space groups \textit{Imma} and \textit{Pnma},
respectively. In order to keep our work transparent, we have
chosen the \textit{Pnma} scheme, preferred in actual works
(\cite{5,6,nijs}). Peculiar details of the UCoGe metallurgy and
crystal structure investigation will be simultaneously published
elsewhere \cite{jirka}. The magnetic measurements were performed
in a commercial SQUID magnetometer (longitudinal geometry) down to
1.8 K and magnetic fields up to 7 T. As a reference an analogous
data series was recorded in a PPMS device using vibration and
extraction magnetometers within the same conditions. To verify the
fully ordered FM state proposed in \cite{5}, we used the a.c.
susceptibility butterfly method \cite{1o}. The polycrystalline
samples were measured as randomly fixed powders; the single
crystal was studied in magnetic fields applied along the principal
crystallographic directions. The electrical resistivity
experiments were carried out in a PPMS device using the $^3$He
option with the current and the external magnetic field applied
parallel to the $c$-axis; other geometries were prevented by the
sample shape. Before each zero-field (ZF) and zero-field-cooled
(ZFC) experiment, respectively, we applied the standard procedure
using the oscillate mode in the SC magnets in order to obtain the
lowest remnant field possible. We reproduced all the low-field
experiments several times in order to eliminate the effect of
remnant field in the SC coils.


\begin{figure}
\scalebox{0.48}{\includegraphics{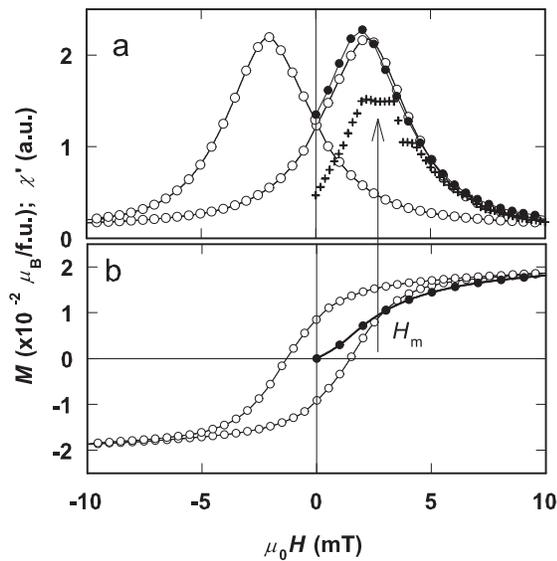}}
 \caption{\label{Fig1}Hysteresis loop and virgin magnetization curve measured
by a.c. susceptibility (a) and common (d.c.) magnetization (b)
measurements at 1.8 K. The panel (a) depicts the butterfly a.c.
loop (open circles) together with the virgin a.c. curve (black
circles). The derivative of the d.c. virgin curve (crosses) almost
coincide with the virgin a.c. curve, and estimates the critical
metamagnetic field, $H_{\mathrm{m}}$.}
 \end{figure}

The crucial point in controlling the SC state by the FM-SF is the
nature of the ZF state above the critical SC temperature. Assuming
a nearly perfect material (from the structural point of view), the
question is whether the UCoGe is a long-range FM or a system with
critical FM-SF. A representative easy-axis ($\mu_0H\parallel c$)
hysteresis loop and a virgin curve recorded at 1.8 K on the
single-crystal is depicted in Figure 1. The Figure 1a shows a
typical a.c. butterfly loop with the two maxima at fields, which
can be attributed to the coercivity field, $H_{\mathrm{c}}$ in the
d.c. loop \cite{1o}. However, the virgin curve does not follow the
maximum-free trend expected for ferromagnets, because the
coercivity and remanence build up after the first field sweep
\cite{1o}. In principle, the maximum can be ascribed to a
crossover between reversible and irreversible dynamic of FM
domains, or a kind of field-induced metamagnetic transition
\cite{1o}. The common d.c. loop in Figure 1b corresponds well to
the data presented by Huy \textit{et al.} \cite{6}, the virgin
curve clearly shows an inflex point almost coinciding with the
maximum on the corresponding a.c. branch.

The static $M$ vs. $\mu_0H$ measurements for fields applied along
the $b$ and $a$ axis, respectively, revealed results in agreement
with \cite{6}. The signal of the corresponding a.c. loops was more
then 10 times lower in comparison to the easy axis response;
however, the maximum was observed at $H\sim H_{\mathrm{m}}$ on the
virgin curve and both branches of the loop, respectively, for both
hard directions. The results of identical experiments performed on
our annealed polycrystalline samples reflect clearly a subdued
character of the easy axis response, suggesting a dominant
character of the $c$-axis contribution (if we ignore any potential
texture, which is expected to be negligible in a randomly-fixed
powder).

In the pilot paper by Huy \textit{et al.} \cite{5}, the formation
of the FM state was proved by appearance of a maximum at around
2.5 K on the temperature dependence of the a.c. susceptibility,
recorded at ZF. In our identical experiments (with the strictly
kept ZF regime), we lacked the maximum for all polycrystalline
samples, however when applying an external magnetic field of 4 mT,
the corresponding maximum was observed. Surprisingly, the
measurements of the a.c. susceptibility (with the a.c. magnetic
field, $H_{\mathrm{ac}}$ = 0.3 mT, and the external d.c. field
applied along the principal axes of the single-crystal) revealed
the $T_{\mathrm{C}}$-related maximum already in ZF for the $a$ and
$b$ axes, respectively. In contrast, the easy axis ($c$-axis)
behavior reflected that of the polycrystalline samples.

The results are shown in Figure 2. In the easy axis, no clear
anomaly occurs in ZF-dependence of the real part of the a.c.
susceptibility ($\chi$'), but in a d.c. magnetic field of 2 mT an
abrupt enhancement of the $c$-axis-related signal together with a
symmetric maximum is observed. When increasing field up to 4 mT
the signal is lowered, the maximum vanishes to re-appear at 6 mT
while the signal becomes continuously suppressed. When increasing
the d.c. magnetic field up to 2 T, the maximum becomes broader and
continuously extends to higher temperatures. If we excite the
system with an a.c. field applied along the $b$-axis (Figure 3b),
the peak appears already in ZF, but is completely inert to the
d.c. field of 2 mT. With increasing the d.c. magnetic field, the
anomaly is smoothly suppressed but peaks almost at the same
temperature. Moreover, the signal at the hard axes is
approximately 10 times lower than that in the easy direction (in
consistency with the butterfly loop experiments). When we focus on
the imaginary part ($\chi$''), we clearly see no anomaly, which
usually demonstrates FM ordering. Although there is a clear
maximum on the $\chi$'-curve for the measurements along the hard
directions, the imaginary part lacks any trend to saturation at
low temperatures. Based on magnetization and a.c. susceptibility
results, we propose a scenario for UCoGe, which considers that in
ZF at temperatures lower than the proposed $T_{\mathrm{C}}\sim$ 3
K a paramagnetic state with strong critical anisotropic FM-SF is
established, contrary to the FM state reported by Huy \textit{et
al.} \cite{5}.

\begin{figure}
\scalebox{0.48}{\includegraphics{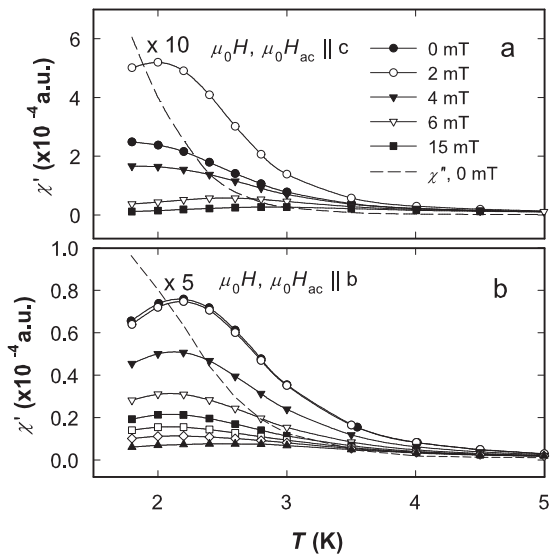}}
\caption{\label{Fig2}Low-temperature a.c. susceptibility (real
part, $\chi$' and imaginary part, $\chi$'') for external d.c.
fields applied parallel to the
 $c$ and $b$ axes. The plots demonstrate the anisotropic behavior of the FM-SF under moderate magnetic fields.}
 \end{figure}

\begin{figure}
\scalebox{0.48}{\includegraphics{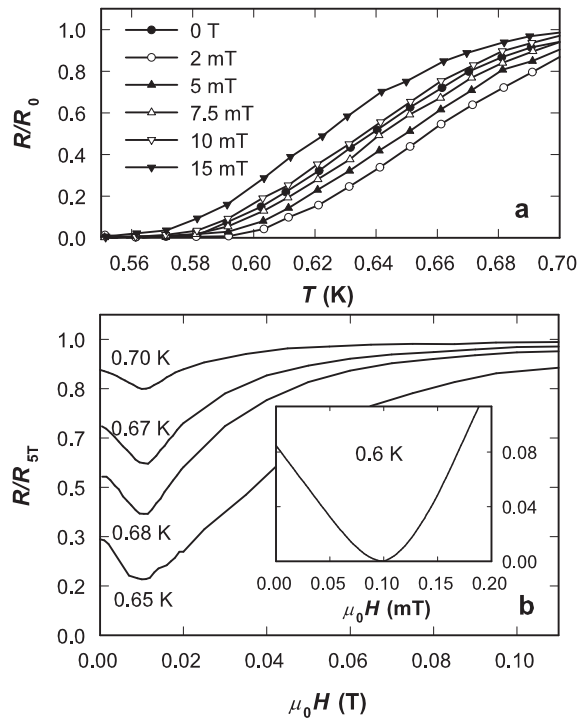}}
 \caption{\label{Fig3}Results of electrical resistivity measurements on the UCoGe
 single crystal ($I\parallel c \parallel \mu_0H$). The panel (a) demonstrates the anomalous evolution of
 the relative electrical resistivity, $R/R_0$ ($R_0$ was obtained from the fit to the normal-state data: A$T^2+R_0$) SC state in the vicinity of the $T_{\mathrm{SC}}$ under various external magnetic fields.
 The panel (b) shows the in-field dependence of the relative electrical resistivity, $R/R_{\mathrm{5T}}$ at temperatures in the vicinity of the $T_{\mathrm{SC}}$.
 The minimum at $\sim$ 0.01 T, attributed to the crossover is enhanced when approaching the homogeneous SC
 state at $\sim$ 0.6 K in our sample (shown in the inset of the Figure 3b).}
\end{figure}

The coexistence of the FM-SF and the USC state was subsequently
explored by detailed measurements of the electrical resistivity
and magnetoresistance (MR) as shown for the single-crystalline
sample in Figure 3. The principal observation is a transient
increase of the $T_{\mathrm{SC}}$ under applied magnetic field.
First, the $T_{\mathrm{SC}}$ (determined as the inflection of the
jump on the $R(T)$ curve) moves from 0.64 K to 0.66 K at $\mu_0H$
= 0 T and $\mu_0H$ = 2 mT, respectively. It further decreases down
to the initial ZF value up to 10 mT, and finally monotonously
decays with increasing magnetic field applied, as expected for a
SC system. The unusual feature is supported by the MR measurements
at temperatures in the vicinity of the $T_{\mathrm{SC}}$ (Figure
3b). The curves exhibit a clear dip in the field of 10 mT, which
becomes continuously smeared out with increasing temperature. The
data recorded on polycrystalline samples are nearly identical.
However, the absolute values of the observed effects are naturally
reduced.

On the basis of the presented experimental results, we propose an
estimative low-temperature phase diagram, depicted in Figure 4
comprising four phases I - IV. There are two essential inquires
about the proposed phase diagram. The first is a competition of
the FM-SF to the robust FM state, already discussed in the part
related to the anisotropic a.c. susceptibility measurements. The
second point is the classification of the two generally different
SC states related to the phase II and III, respectively. Strictly
speaking, the major controversy points to the nature of the
low-field SC state related to the phase II.

\begin{figure}
\scalebox{0.45}{\includegraphics{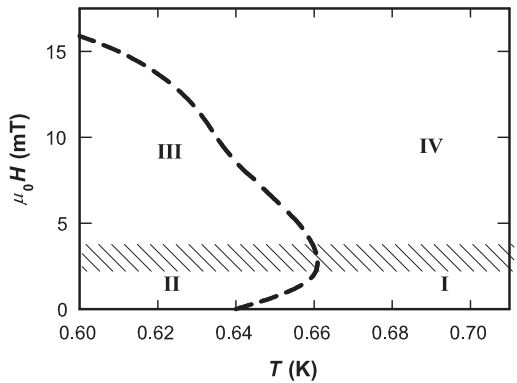}}
 \caption{\label{Fig4}The detail of the tentative FM-SC
 phase diagram. The phase I corresponds to the state below the critical field $H_{\mathrm{m}}$, characterized with FM-SF.
 The phase II represents the SC state with the anomalous evolution of the
 $T_{\mathrm{SC}}$. The phase III is attributed to the previously reported USC
 state with the proposed $p$-wave SC state coexisting with FM, and the phase IV is ascribed to the FM state,
 which is provoked by applied magnetic field from the FM-SF.}
 \end{figure}

Let us propose two fundamentally diverse scenarios. In the first
path, the two SC states are of different symmetries. In ZF, the
order parameter is supposed to be isotropic (due to expected easy
axis disorder, which might prevent stable zero-field FM) resulting
in the only allowed $s$-wave SC \cite{gorkov}, but when we reach
the robust FM state, the anisotropy of the Fermi surface is fully
developed and the unconventional $p$-wave spin-triplet state can
be formed. The proposed mechanism can be reflected in theories of
Blagoev and Fay \cite{1t,fay}, which claim, that for weakly FM
metals the coexistence of longitudinal SF and a gapless Goldstone
mode allows formation of both the $s$-wave (generalized FFLO
state) and the $p$-wave SC. The crucial discrepancy is, that the
$p$-wave SC state should occur on the paramagnetic limit and the
$s$-wave SC on the FM. However, the crossover in the anisotropy
and spin dynamics of the FM-SF in the vicinity of the potential
QCP do not exclude simultaneous occurrence of the two SC states
with a generally different symmetries of the SC-OP. On the other
hand, we can consider the effect of FM domains in USC, so the SC
phases have the same symmetry of the OP. For an orthorhombic
system with uniaxial FM, four symmetry classes are allowed, which
can be split in two pairs comprising two equivalent
co-representations; in general, each of them may have a different
$T_{\mathrm{SC}}$. Within one of the pairs, the sub-states in
magnetic domains with the magnetic moments oriented parallel or
antiparallel to the magnetic field can be expressed by the two
co-representations. Within the above-described picture, either a
crossover between the two sub-states of the two
co-representations, or between the two pairs of states may occur.
In general, there is no straightforward proof for the competition
of the two analogous pairs of the SC classes in the vicinity of
QCP. Therefore we propose, that the anomalous behavior of the
$T_{\mathrm{SC}}$ in phase II is probably not due to the crossover
between the two different pairs of classes, but the applied
magnetic field rather influences the critical FM-SF as the
governing factor conditioning the phase line of the SC state.
Considering the previous scenario of ZF disorder, the magnetic
field stabilizes the FM-SF, and the proper $p$-wave spin-triplet
SC state is established within one of the pairs of the two allowed
symmetry classes.

In conclusion, we have investigated low-field phenomena in UCoGe
stimulated by the interplay of FM-SF and superconductivity in the
vicinity of the FM-QCP. We observed two unique features: 1.
unusual enhancement of $T_{\mathrm{SC}}$ under applied magnetic
fields, and 2. evidence of FM-SF in zero magnetic field, contrary
to the claimed zero-field FM order. We constructed a tentative
phase diagram considering the critical phenomena on the border of
the SC region. Finally, our observations propose a crossover
between two SC phases, with potentially different symmetries of
the OP. However, the nature of the phase II is so far speculative.
Fine experiment, typically angle-dependent in-field measurements
on a well-defined single-crystal, in order to determine the
corresponding symmetry of the SC gap in the II and III phases,
respectively, would help to discriminate between the
above-discussed scenarios.

This work is a part of the research plan MSM 0021620834 that is
financed by the Ministry of Education of the Czech Republic.

\end{document}